\documentclass[MSNbibl,nameyear,dvips]{arxstspdf}
\usepackage{flushend}
\usepackage{stfloats}

\volume{26}
\issue{2}
\pubyear{2011}
\firstpage{175}
\lastpage{178}
\doi{10.1214/10-STS318B}
\referstodoi{10.1214/10-STS318}

\begin{document}
\begin{frontmatter}

\title{Discussion of ``Calibrated Bayes, for Statistics in General, and Missing Data in
Particular'' by R. J. A. Little}
\runtitle{Discussion}
\pdftitle{Discussion of Calibrated Bayes, for Statistics in General, and Missing Data in
Particular by R. J. A. Little}

\begin{aug}
\author[a]{\fnms{Michael D.} \snm{Larsen}\corref{}\ead[label=e1]{mlarsen@bsc.gwu.edu}}
\runauthor{M. D. Larsen}

\affiliation{George Washington University}

\address[a]{Michael D. Larsen is Associate Professor of Statistics, Biostatistics Center,
George Washington University, 6110 Executive Boulevard, Suite 750 Rockville, Maryland~20852, USA \printead{e1}.}

\end{aug}



\end{frontmatter}

I would like to thank Rod Little for a thought-provoking and
well-presented paper on the ``calibrated Bayes'' approach to statistics.
The author makes a~strong case for the advantages of Bayesian methods
and multiple imputation when dealing with missing data: the ability to
fill in the data while accounting for the missing information in the
inference is highly desirable. The article expounds the idea of a
calibrated Bayesian approach to statistical problems in general and to
missing data issues in particular. It would certainly be interesting to
see an expanded treatment of how to implement calibration in the
Bayesian context. Does this primarily mean selecting and transforming
variables and models to get a good fit to the data? Does it also mean
running more analyses to check sensitivity to missing data and
model/variable assumptions? What about hierarchical models (e.g.,
Bayarri and Castellanos, \citeyear{BayCas07})? Advances in (MCMC) algorithms, computing
power and (free) software on the web have made Bayesian approaches
feasible for a~much broader group of statisticians and other
researchers. Indeed, a~significant portion of the article summarizes and
illustrates some techniques. There is a need for more ``how to be
calibrated'' guidance, including computing tools and textbook examples,
for applied Bayesians in practice.

One example from recent work comes to mind. In this example, a
frequentist analysis is going to be reported, but there are missing
data. Multiple imputation in this context is useful for building
confidence in the results, because it is possible to compare and
contrast results under different missing data assumptions. In an
additional analysis of data from the Diabetes Prevention Program
(Knowler et~al., \citeyear{Knoetal02}), parent's age at death was being used as a
predictor of the onset of diabetes in a population of adult
pre-diabetics. Parents who live a~long time generally are a good
predictor of health of their children; the premature death of a parent
does not augur well for offspring. But nearly 1$/$3 of the parents were
still alive at the beginning of the study (when parental age at death
was captured). Not surprisingly, these parents were less likely to have
had a~cardiovascular event in the past than were the other parents.
Their adult children tended to be younger than the other study
participants. In the analysis using parent's age at death as a predictor
variable, should data from the 1$/$3 of the subjects be discarded from the
analysis?

An attempt was made to model time until death for the parents who were
living at study entry. Several variables were predictive of parental
longevity. It was, then, possible to multiply impute age at death under
some models, and then conduct the primary analysis utilizing multiple
imputation combining rules. In the end, the results did not change much
from the analysis based on only the complete cases---other than being
younger, the patients with living parents did not differ much on average
from the others. Even if a Bayesian analysis is not ultimately reported
in detail, use of a~multiple imputation procedure did seem to lead
credence to the frequentist-procedure results; that fact can be stated
very succinctly in a medical journal article. Statistical practice would
move closer to ``calibrated Bayes'' if checks such as the one described
here became standard and expected instead of novel.

If the analysis in the example described above had been substantially
different from the complete case answer, then more work (i.e.,
statistical modeling and model checking on the available data) would
have been needed to understand why. One might then discover something
important in the data that would not be apparent for either analysis
alone. Today, one could imagine that substantially more effort would
have been needed to get an alternative Bayesian analysis accepted in
many journals as the primary analysis instead of the complete case
analysis. Statistics in practice would be closer to ``calibrated Bayes''
if well done Bayesian analyses were more likely to be put forward as the
primary analyses. Indeed, as the author notes, more work is needed in
the area of diagnostics for the quality of multiple imputations (Su
et al., \citeyear{autokey14}), which has implications for the acceptability of
MI analyses.

Let me make three additional comments, two brief and one not as brief.
First, the author states, accurately, that it is now easier than before
to implement Bayesian analyses and multiple imputation. Still, there is
need to have applied statisticians who understand computational details.
For example, the author does not mention how to get standard errors (in
Example 2) from maximum likelihood when there are missing data. There
are ways to do this, of course, but are they easily accessible in
current computational tool kits? Also, efficient computation and
efficient algorithms are still important. Computing time is still a
factor that limits many studies. Evolving options for computing in large
problems should enter the mainstream of applied statistical practice and
thereby facilitate the implementation of calibrated Bayesian analyses.
This is not to say that frequentists do not encounter computational
issues. Indeed, simulation and bootstrap are important tools for
studying behavior of procedures in small- or moderate-sample size
situations under null and alternative models.

Second, the author mentions sequential regression multivariate
imputation (SRMI), also referred to as multiple imputation through
chained equations\break (MICE), and penalized spline of propensity prediction
(PSPP) as two alternatives to simpler models. The latter the author
argues has a double robustness quality: if either the prediction model
or the response propensity model is correct, then the estimator based on
the imputed data is consistent. In survey sampling, donor-based
procedures referred to as ``hot deck'' procedures are often used to fill
in missing values. Good hot deck procedures use matching information in
manners similar to multivariate matching or propensity matching to pick
similar donors. Donors have observed values that are real and consistent
with true association patterns in the data set and with dependencies
among variables that are challenging to model. Well-designed multiple
imputation hot deck approaches and mixes of hot deck and modeling
approaches could provide an alternative, that could be acceptable to
statisticians of both Bayesian and frequentist persuasions, to MI
approaches.

Third, in Section 2 the author divides the statistics world into
frequentists and Bayesians. This division is clearly the focus of the
paper and useful for the discussion, but a broader view is possible.
There are statisticians who think of themselves as survey samplers; both
the author and discussant have connections to this world. Survey
samplers follow procedures as described in textbooks such as those by
Cochran (\citeyear{Coc77}) and S\"{a}rndal, Swensson and Wretman (\citeyear{SarSweWre92}) for making
inference about finite population values. The randomness in these
procedures comes from (controlled) random selection from a finite
population of units. It is related to frequentist inference, but the
``parameters'' can look different, for example, $\bar{y} = \sum_{i=1}^N
y_i/N$, the finite population average instead of $\mu$. Generally the
goal in survey inference is frequentist in nature: 95\% confidence
intervals based on probability samples from the current sample frame
should contain their target population quantity at least 95\% of the
time and not be wider than necessary. The stated goal implies that 95\%
coverage should occur as well in samples from conceptually similar
sample frames.

In large scale surveys, additional steps often are taken, such as coding
and editing survey responses, forming post strata, and survey weight
adjustment, that are not clearly motivated by frequentist principles
aimed at estimating a model parameter $\theta$. Forming post strata aims
at reducing variance and also bias in surveys with nonresponse. Survey
weight adjustment can have other goals, including matching published
population totals and other published results. In fact, in survey
sampling there is a method of adjusting survey sampling weights called
``calibration weighting'' or ``calibration estimation'' (Deville and
S\"{a}rndal, \citeyear{DevSar92}). This method brings weighted estimates from a sample
in line with (possibly several) published totals, thereby making all
estimates using the adjusted weights potentially more relevant to the
finite population.

Most survey sampling textbooks do not even mention Bayesian ideas, as
the heyday of (now) classic textbooks in survey sampling (1950s, 1960s)
was\vadjust{\eject} definitely not a time of Bayes popularity. Further, an appealing
aspect of randomization inference in survey sampling is that no model is
involved at all. Of course, that does not mean that all survey
inference procedures are advisable independent of context. In survey
sampling, ratio estimation is a standard choice. But it only works well
if there is a reasonably strong positive correlation between an outcome
and an auxiliary variable. The model that is consistent with such an
approach actually is a linear model with zero intercept, which is a
rather restrictive model. In general, understanding model limitations
(or implied model limitations) should help in picking good estimators.
Being ``calibrated'' in the sense of Little's article surely would include
checking the fit of the implied models to survey data and consideration
of broader modeling options such as those described in S\"{a}rndal,
Swensson and Wretman (\citeyear{SarSweWre92}) along with calibration estimation ideas.
Checking the fit is important for avoiding inconsistencies between
models and data. It also could be part of efforts to improve efficiency
in estimation.\looseness=1

One textbook reference that does consider survey sampling from a modern
Bayesian perspective is Section 7.4 of Gelman et al. (\citeyear{Geletal04}); see
also references in Section 7.10. There also are a number of relevant
references in the literature. Gelman (\citeyear{Gel07}) compared survey weighting,
regression modeling and related Bayesian approaches. Little (\citeyear{Lit93})
discussed modeling as related to post stratification. Techniques of
small area estimation (e.g., Rao, \citeyear{Rao03}, and Jiang and Lahiri, \citeyear{JiaLah06}) have
utilized hierarchical models along with various approaches to
estimation. Lu and Larsen (\citeyear{LuLar07}, \citeyear{LuLar08}) used hierarchical modeling with
model selection in a finite population survey application.

The connection between Bayesian methods in general and finite population
survey sampling will need more elaboration and development before a
Bayesian analysis is accepted by the majority of survey researchers. The
use of multiple imputation for missing data in the survey context,
though, should not have the same high hurdle to cross. Multiple
imputation and small area estimation likely will be techniques that lead
survey samplers toward a calibrated Bayesian approach. Once a data set
has been adequately imputed there is no reason not to use survey weights
and survey estimators. In fact, one could use survey calibration
weighting on multiply imputed data sets. The key issue is how to
determine if a data set has been adequately imputed. Gelman (\citeyear{Gel10})
quotes Hal Stern when noting that perhaps the largest divide is between
those who model and those who do not model the data. One can question
the choice of a parametric model, or likelihood function, and the
specification of a prior distribution. Concerns about being consistent
with the data versus the goal of extracting information from the data
through models could be the real source of division between approaches.
Flexible modeling options incorporated into multiple imputation methods
(e.g., MICE, SRMI and the author's PSPP) aim specifically to address
consistency concerns while enabling multiple imputation. It remains to
present results and diagnostics in a convincing manner. Reporting
diagnostic checks on consistency and acknowledging model limitations in
a Bayesian analysis could have advantages in terms of helping establish
credibility in a wider community.

Besides the frequentists, Bayesians and survey samplers, there is a
substantial group of applied researchers who use statistical methods
primarily because they are the standard procedures in their fields and
encoded in familiar statistical software. Usually these are frequentist
procedures that involve estimating parameters, but, like nonparametric
methods, they do not have to be. Some classification and discrimination
procedures, for example, do not have clearly identifiable parameters
that are estimated. Indeed, classification trees grow with the available
data and the main output is a measure of (cross-validation) accuracy.
How does this relate to the ideas of this article? Surely there is a
sense in which any method in use for analyzing data should be calibrated
to reality, whether that reality is expressed in terms of probability
distributions and their parameters, finite population characteristics or
replicable experience. If software makes more Bayesian methods readily
available and guidance and experience makes them acceptable (even
preferable) and well known, then calibrated Bayesianism will have wider
reach into statistical practice.

\vspace*{2pt}
\section*{Acknowledgments}
\vspace*{2pt}

Again, I would like to thank the author for
his stimulating article. I would also like to thank the Editor for
helpful comments on the discussion. Hopefully the article motivates the
development of practical methods and advice for following the path of
calibrated Bayesian statistics.


\end{document}